\RequirePackage{fix-cm}

\documentclass[natbib,smallextended]{svjour3}       
\smartqed  
\usepackage{graphicx}
\usepackage{amsmath}
\usepackage{graphicx,psfrag,epsf}
\usepackage{enumerate}
\usepackage{soul}
\usepackage{natbib}
\usepackage{ulem}
\RequirePackage{listings}
\RequirePackage[linesnumbered,ruled]{algorithm2e}
\usepackage[usenames, dvipsnames]{color}
\RequirePackage{amssymb}
\usepackage{mathtools}
\RequirePackage{bm}
\usepackage{algorithmic}
\usepackage{caption}
\newenvironment{step}[1][htb]
  {
   \begin{algorithm}[#1]%
  }{\end{algorithm}}
\newenvironment{summary}[1][htb]
  {
   \begin{algorithm}[#1]%
  }{\end{algorithm}}

\begin{document}

\title{T-optimal  designs for multi-factor polynomial regression models via a semidefinite relaxation method
}

\titlerunning{T-optimal design for multi-factor polynomial models}        

\author{Yuguang Yue         \and
        Lieven Vandenberghe \and
        Weng Kee Wong 
}


\institute{Yuguang Yue \at
              Department of Biostatistics, University of California, Los Angeles\\
              \email{yuguang@utexas.edu}           
           \and
           Lieven Vandenberghe \at
              Department of Electrical Engineering, University of California, Los Angeles\\
              \and
              Weng Kee Wong \at
              Department of Biostatistics, University of California, Los Angeles
}

\date{Received: date / Accepted: date}

\maketitle

\begin{abstract}
We consider T-optimal experiment design problems for discriminating
multi-factor polynomial regression models where the design space is
defined by polynomial inequalities and the regression parameters
are constrained to given convex sets.
Our proposed optimality criterion is formulated as
a convex optimization problem with a moment cone constraint.
When the regression models have one factor, an exact
semidefinite representation of the moment cone constraint can be applied
to obtain an equivalent semidefinite program.
When there are two or more factors in the models, we apply a moment relaxation technique
and approximate the moment cone constraint by a hierarchy of
semidefinite-representable outer approximations.
When the relaxation hierarchy converges, an optimal
discrimination  design can be recovered from the optimal moment matrix,
and its optimality is confirmed by an equivalence theorem.
The methodology is illustrated with several examples.
\keywords{Continuous design\and Convex optimization\and Equivalence theorem\and Moment relaxation\and Semidefinite programming}
\end{abstract}

\section{Introduction}
\label{sec:intro}

In many scientific investigations, the underlying statistical model that
drives the  outcome of interest is not known.
In practice, researchers may be able to identify a few plausible models
for their problem and an early goal is to find a design to collect data optimally to identify the most appropriate model.  Once this task is accomplished, one proceeds to the next phase of the scientific investigation, which may be to estimate parameters in the selected model or use the model for making statistical inferences, such as predicting values of the responses at selected regions. Alternatively, one performs model diagnostics after the data are collected and evaluates whether the model assumptions are valid. We propose a new method for finding  an optimal design to
discriminate  among several multi-factor polynomial regression models
defined on a user-selected compact multi-factor design space $\mathcal{X}$ and show that it is straightforward to implement our strategy.

\cite{atkinson1975design,atkinson1975b} were among the first to formulate a statistical framework for finding an optimal discrimination design when the class of plausible models is defined on a user-defined space.  The response variable is univariate and continuous,  and all errors are  assumed to be normally distributed, each with mean zero and equal variance.  This has been the traditional setup for finding optimal discrimination designs until recently where errors are allowed to be non-normally distributed.
%


Our work assumes all models of interest are polynomial models where the cone of possible moment matrices  may be represented as an exact semidefinite representation.  This is possible when the polynomial model has one factor, but not for general polynomial models \citep{shohat1943problem,scheiderer2018spectrahedral}.
 For the latter situation, we use the moment-sum-of-square hierarchy to
approximate the moment cone.

 Our method has several advantages over current methods for discriminating several models. First, we do not need to assume a known true (null) model among the plausible models.  Until recently, this assumption is required in the optimal discrimination design and is a frequent critique of the setup, see for example, \cite{fedorov1972optimal,atkinson1974planning,atkinson1975design,atkinson1975b} and \cite{duarte2015semi}; our setup permits possible values of the parameters to belong to any convex sets $\Theta_{j}$ and not singleton sets. Second, unlike some of the state-of-the-art algorithms that require the design space be discretized for finding an optimal design \citep{yang2013optimal},  our method does not require us to replace a continuous design space by a set of candidate design points.  This is an important consideration because for high dimensional problems where we have several factors, methods based on discretizing the search space are likely to be slow.    A third advantage of our method is that we do not require  the number of support points  in the optimal design to be specified in advance.  This is in contrast to several mathematical programming approaches, such as those in \cite{duarte2015semi}, where the semi-infinite programming algorithm requires a pre-selected number of design points to start with.
Fourth, our approach is flexible in that the design space can be defined by polynomial inequalities to more realistically capture the physical or cost constraints of the design problem.  For example in mixture models, where the mean response is commonly modeled using Scheff\'e's, Becker's or Kasatkin's polynomial models \citep{wong}, our framework can directly incorporate polynomial constraints on the design space.

Section 2 describes the statistical background of experiment designs. Section 3 provides the formulation of the optimal discrimination design problem along with the exact representation of a special case, and introduces the hierarchy moment relaxation algorithm that includes moment relaxation theory, equivalence theorem and the solution extraction method.  Section 4 shows the detailed steps of our algorithms. In Section 5, we provide six examples, some of which are specially selected  to demonstrate that the algorithm generates the same theoretical optimal designs in the literature.    Section 6  provides a summary and a brief discussion on some of the unpredictable properties of the optimal discrimination designs, some limitations of our approach and future direction of our work.  In the appendix, we provide a sample Matlab code that we used  to generate one of the optimal discrimination designs in this paper.

\section{Background}
\label{sec:backgnd}
An experimental design is optimal if it optimizes a given criterion over the set of all designs on the design space.  The  most common design criterion is D-optimality  that seeks to minimize the determinant of the covariance matrix of the estimated parameters \citep{fedorovbook,waterhouse2009optimal,duarte2018adaptive}. Much less attention has been given to finding an optimal design that discriminates among competing models. The theoretical framework for experimental design for model discrimination using T-optimality was established in a series of papers by \cite{fedorov1972optimal,atkinson1974planning} and \cite{atkinson1975design,atkinson1975b}. The typical setup assumes that we want to discriminate between two parametric models, one of which is a fully known parameterized `true model' and the other is a `test model' with a known mean function apart for the values of the parameters. The T-optimal design maximizes the lack of fit sum of squares for the second model by maximizing the minimal lack of fit sum of squares arising from a set of plausible values of the unknown parameters \citep{fedorov1972optimal,atkinson1975design}.

 Additional theoretical developments can be found in \cite{de1991optimum,dette1994discrimination,fedorov2012model,wiens2009robust} and, \cite{dette2009optimal}. \cite{ucinski2005t} proposed a generalized criterion for multi-response model and \cite{carlos1993optimum} gave a criterion for discriminating between binary outcome models. T-optimality has been applied to discriminate among  polynomial models \citep{dette2012t},  Fourier regression models \citep{dette2003optimal}, Michaelis–Menten kinetic models \citep{atkinson2012optimum} and dynamic systems described by sets of ordinary differential equations \citep{ucinski2004optimal}.When errors are not normally distributed, KL-optimality is used instead of T-optimality to discriminate  among models; see \cite{lopez2007optimal}, among others.  Most recently, \cite{melas} proposed an interesting and relatively easy method to find an optimal design to discriminate among semi-parametric models.

In general, finding T-optimal designs is a challenging problem especially when multiple models are involved
because the design criterion is not differentiable and the structure of the optimization problem has two or more layers of nested optimization.   Consequently,  formulae for T-optimal designs are only available for simple optimal discrimination design 
problems. Several algorithms have been proposed to specially find a T-optimal design. 
Some examples are \cite{wynn1970sequential, fedorov1971design,fedorov1972optimal,atkinson1975design,atkinson1975b} where they
sequentially add one or more points specially selected to the current design.    Unrelated to previous methods, \cite{duarte2015semi} proposed using semi-infinite programming to solve an optimal discrimination design problem. 

We focus on finding an optimal discrimination design under a concave
criterion when we have a polynomial regression model with several factors on
a compact, possibly non-convex, design space $\mathcal{X}$.
We denote the mean  response for each of the $K$ possibilities by $\eta_1(x)$, \ldots, $\eta_K(x)$ and
let $f(x)$ be the $l\times 1$-vector of basis monomial functions of regressors
with
input factor $x$. The method is not restricted to monomials and can be readily extended to other polynomial
basis functions.

The mean response function for each of the $K$ models is
$$\eta_j(x) = \theta_j^Tf(x) + \epsilon_j, \quad j=1, \ldots, K,$$
where $\theta_j^T = [\theta_{j1},\ldots,\theta_{jl}]^T$ is a partially unknown
parameter vector
and $\epsilon_1,\ldots, \epsilon_K$ are independent random noises, each with standard Gaussian distribution with zero mean and unit
variance.
We assume that for the $j^{th}$ model, there is a known set $\Theta_j$ that contains all
the possible values of the parameter vector $\theta_j, j=1,\ldots,K$, and all such sets are distinguishable.
The constraint $\theta_j \in \Theta_j$ may include the constraint that
certain coefficients of $\theta_j$ are zero, i.e., the  model $j$ does
not involve certain basis functions in $f(x)$.  This allows us
to simplify our notation and use the same vector of basis functions
$f(x)$ for each model.

Suppose we have resources to take $N$ observations for our study and $\xi$   is a design that takes $p_i$ proportion of the $N$ observations at the design point $x_i\in \mathcal{X}$, $i=1,\ldots,m$.  We call such a design a continuous design and represent it by
$$
\xi =
 \begin{pmatrix}
  x_1 & \cdots & x_m \\
  p_1 & \cdots & p_m
 \end{pmatrix}.
$$
The first row displays the support points of the design and the second row displays the proportion of observations to be taken at each of the support points. Clearly $p_1+\ldots+p_m=1$.  In practice, the continuous design is implemented by first rounding each  $Np_i$ to an integer  subject to the constraint that they sum to $N$.
The advantages of  working with continuous designs are  that , when the criterion is concave, a unified theory exists for finding  optimal continuous  designs and there are algorithms for finding a variety of optimal designs. Further, there are equivalence theorems to confirm optimality of a design and simple analytical tools for checking proximity of a design to the optimum without knowing the optimum.

\section{Solution via semidefinite optimization}
\label{sec:algo}
\subsection{Reformulation of the design problem}
A $T$-optimal design  maximizes the minimal squared distance between the mean responses from two
possible polynomial models.
 Each of our regression models is defined on $\mathcal{X}$ and can be represented as
$$\eta_j(x)=\theta_j^Tf(x) + \epsilon_j, \quad j = 1, 2,...K$$
$\theta_j \in\Theta_j, \epsilon_j\sim N(0,\sigma^2)$.
The vector $f(x)$  is $l$-dimensional and  contains the $l=C_n^{n+d}= (n+d)!/(n!d!)$
monomials of degree $d$ or less in the $n$ factors $x$.

We recall the $l\times l$   matrix
$$ M(\xi)=\int_{\mathcal{X}}f(x)f(x)^Td\xi $$
is the moment matrix of the design $\xi$ and $M(\xi)$ is a linear function
of $\xi$.  In what is to follow, we write  $M(\xi)$ as $M$ and optimize it as a variable directly under a given criterion.  The sought design   is then recovered from the optimal $M$.

Given a design $\xi$, let $\Delta_{jk}(M)$ be the non-centrality parameter for the
pair of models $j$ and $k$ defined by
\begin{eqnarray*}
\Delta_{jk}(M)
& = & \inf_{\theta_j\in\Theta_j, \theta_k\in\Theta_k}
\int(\theta_j^Tf(x)-\theta_k^Tf(x))^2 d\xi\\
& = & \inf_{\theta_j\in\Theta_j, \theta_k\in\Theta_k}
(\theta_k-\theta_j)^TM(\theta_k-\theta_j).
\end{eqnarray*}

We note that the function $\Delta_{jk}(M)$ is concave in $M$
because it is the pointwise infimum of a family of linear functions of
$M$ \citep{boyd2004convex}.   
The quantity $\Delta_{jk}$ represents the infimum of the squared distance between the mean responses from
model $j$ and model $k$ over all possible choices of the parameters
in $\Theta_j$ and $\Theta_k$.
If we wish to discriminate between model $j$ and model $k$ only, the T-optimal design maximizes
$\Delta_{jk}(M)$ over all designs $\xi$.
If there are $K > 2$ models  to discriminate, there are $K(K-1)/2$ such criteria, one for
each pair of models. These are generalizations of the non-centrality parameters discussed in \cite{atkinson1975design, atkinson1975b}, where they noted that T-optimality applies to nested models only when there are constraints placed on the model parameters; otherwise the non-centrality parameter is zero and the T-optimality criterion is not appropriate. \cite{atkinson1975design} provides an  illustrative example of such a situation; see also the introduction in \cite{dette2009optimal} where they provided a motivation for T-optimality.

Our optimal discrimination design problem involving several models can be reduced to a single-objective optimization problem
by considering
\begin{equation}\label{eq:2}
\begin{aligned}
& \underset{\xi}{\text{maximize}} & &  \min_{j>k}\Delta_{jk}(M) \\
& \mbox{subject to} & & M\in\mathcal{M}
\end{aligned}
\end{equation}
where $\mathcal{M}$ is the set of possible moment matrices $M$.
Alternatively, we can maximize a weighted sum, as in
\begin{equation}\label{eq:3}
\begin{aligned}
& \underset{\xi}{\text{maximize}} &&  \sum_{j>k}w_{jk}\Delta_{jk}(M) \\
& \mbox{subject to} && M\in\mathcal{M},
\end{aligned}
\end{equation}
with nonnegative weights $w_{jk}$. Let
$$\delta_{\Theta}(x) = \begin{cases}
               0 & x\in\Theta\\
               +\infty& \text{otherwise}\\
            \end{cases}$$
  be the indicator
function of the closed and convex set $\Theta$ with a nonempty relative interior.  We then use convex duality theory and derive an alternative expression
for $\Delta_{jk}(M)$.  The function $\Delta_{jk}(M)$
is the optimal value of the convex optimization problem
\begin{equation}\label{eq:444}
\begin{aligned}
& \underset{u,\theta_j,\theta_k}{\text{minimize}}
& &  u^TMu + \delta_{\Theta_j}(\theta_j) + \delta_{\Theta_k}(\theta_k) \\
& \mbox{subject to} & & u = \theta_j-\theta_k
\end{aligned}
\end{equation}
with variables $u$, $\theta_j$, $\theta_k$.
The Lagrangian for this problem  is
\[
L(u,\theta_j,\theta_k) =
 u^TMu + \delta_{\Theta_j}(\theta_j) + \delta_{\Theta_k}(\theta_k)
 + z^T ( u - \theta_j +\theta_k)
\]
and the dual function is the unconstrained infimum of $L$ over
$u$, $\theta_j$, $\theta_k$:
\begin{eqnarray*}
\inf_{u,\theta_j,\theta_k}L(u,\theta_j,\theta_k)
& = & \inf_u{(u^TMu + z^Tu)}
  + \inf_{\theta_j} (\delta_{\Theta_j}(\theta_j) - z^T\theta_j) +\\
&  & \inf_{\theta_k} (\delta_{\Theta_k}(\theta_k) + z^T\theta_k) \\
& = & -\frac{1}{4} z^T M^{-1}z
  - \delta_{\Theta_j}^*(z)
  - \delta_{\Theta_k}^*(-z),
\end{eqnarray*}
where $\delta_{\Theta}^*(x)$ is the value of the support function of the set $\Theta$ at the point $x$.
We recall from \cite{rockafellar1970convex} that the support function of a set $C$
is defined as the conjugate of the indicator function, i.e.
\[
 \delta^*_C(y) = \sup_{x}{(y^Tx - \delta_C(x))}=
\sup_{x\in C}{(y^Tx)}.
\]
  In the minimization of the Lagrangian
we have assumed that $M$ is invertible.   If $M$ is not invertible,
the first term $z^TM^{-1}z$ in the expression for the dual
function should be interpreted as $+\infty$
if $z$ is not in the range of $M$, and as $z^TM^+z$ otherwise,
where $M^+$ is the pseudo-inverse of $M$.

Accordingly, the dual problem is
\begin{equation}\label{eva}
\begin{aligned}
&\underset{z}{\text{maximize}}
& &  -\frac{1}{4}z^TM^{+}z-\delta^*_{\Theta_j}(z) -\delta^*_{\Theta_k}(-z)
\end{aligned}
\end{equation}
with variable $z$, which can be further rewritten as
\begin{equation}\label{eva-2}
\begin{aligned}
&\underset{z, t}{\text{maximize}} & & {-t} -\delta^*_{\Theta_j}(z) -\delta^*_{\Theta_k}(-z) \\
& \text{subject to} &&
 \left[\begin{array}{cc} M & z \\ z^T & 4t \end{array}\right]
 \succeq 0,
\end{aligned}
\end{equation}
with variables $t$ and $z$.
From convex duality theory, the
problems~(\ref{eq:444}) and~(\ref{eva-2}) have the same optimal
values.  It follows that for fixed $M$ the optimal value of~(\ref{eva-2}) is also
$\Delta_{jk}(M)$.
However since the constraint is convex in $M$, we can
jointly optimize over $M$, $t$, and $z$ to maximize $\Delta_{jk}(M)$.
This observation allows us to write problem (\ref{eq:2}) as
\begin{equation}
\label{e-dual1}
\begin{aligned}
& \underset{M, t_{jk}, z_{jk}}{\text{maximize}}
& & \min_{j>k} (-t_{jk})
    - \delta_{\Theta_j}^*(z_{jk}) - \delta_{\Theta_k}^*(-z_{jk}) \\
& \text{subject to}
&&
 \left[\begin{array}{cc} M & z_{jk} \\ z_{jk}^T & 4t_{jk}
 \end{array}\right] \succeq 0, \quad j>k \\
&&& M\in\mathcal{M},
\end{aligned}
\end{equation}
with variables $M$, $t_{jk}$ and $z_{jk}$ for $j<k$.
A similar reformulation for problem (\ref{eq:3}) leads to
\begin{equation}
\label{e-dual2}
\begin{aligned}
& \underset{M, t_{jk}, z_{jk}}{\text{maximize}}
& & -\sum_{j>k} w_{jk} t_{jk}
    - \delta_{\Theta_j}^*(z_{jk}) - \delta_{\Theta_k}^*(-z_{jk}) \\
& \text{subject to}
&&
 \left[\begin{array}{cc} M & z_{jk} \\ z_{jk}^T & 4t_{jk}
 \end{array}\right] \succeq 0, \quad j>k \\
&&& M\in\mathcal{M}.
\end{aligned}
\end{equation}

In practice, the sets $\Theta_j,j=1,\ldots,K$ are often simple, and their
support functions are easy to compute.
The main difficulty in solving~(\ref{e-dual1}) and~(\ref{e-dual2})
is in the moment constraint $M\in\mathcal M$, which is a set that is hard to characterize efficiently even though it is a convex set.  The set $M\in\mathcal M$ can only be described analytically in
 a few special cases.  For example, for univariate polynomials,
the set $\mathcal{M}$ can be represented exactly by a set of semidefinite constraints, using classical results
from moment theory.  Consider, for example,
$\mathcal X = [0,1]$ and $f(x) = (1, x, \ldots, x^d)$, the set $\mathcal M$ is defined by
two linear matrix inequalities
\[
M = \left[\begin{array}{ccccc}
 1 & c_1 & c_2 & \cdots & c_d \\
 c_1 & c_2 & c_3 & \cdots & c_{d+1} \\
 c_3 & c_4 & c_5 & \cdots & c_{d+2} \\
 \vdots & \vdots & \vdots& & \vdots \\
 c_{d-1} & c_{d}  & c_{d+1}  & \cdots & c_{2d-1}
\end{array}\right] \succeq 0
\]
and
\[
\left[\begin{array}{cccc}
 c_1-c_2 & c_2-c_3 & \cdots & c_{d}-c_{d+1} \\
 c_2-c_3 & c_3-c_4 & \cdots & c_{d+1}-c_{d+2} \\
 \vdots & \vdots& & \vdots \\
 c_{d} - c_{d+1}  & c_{d+1} - c_{d+2}  & \cdots & c_{2d-1} -c_{2d}
\end{array}\right] \succeq 0,
\]
see, for example, Theorem 1.1 in \cite{karlintchebycheff}.
In other cases, one can resort to approximating $\mathcal M$
by outer semidefinite approximations
using techniques that have been developed recently in
semidefinite programming methods for polynomial optimization, see for example, \cite{lasserre2001global,lasserre2015introduction}.
This is explained in more detail in the next section.

\subsection{Moment relaxation}
Except in special cases like the one discussed above, \cite{scheiderer2018spectrahedral} has proved that the moment cone is not semidefinite representable, i.e. it cannot be expressed as the projection of a linear section of the cone of positive semidefinite matrices. An important tool commonly used to find optimal designs for polynomial regression is the theory of moment relaxation introduced by \cite{lasserre2001global,lasserre2009moments}. In what is to follow, we  give a brief heuristic introduction of this concept. In particular, we  illustrate how the complicated moment matrix constraint can be represented as semidefinite constraints after we introduce some additional and necessary notation.

\subsubsection{Truncated moment cone}

Let $x^{\alpha}:=x_1^{\alpha_1}\cdots x_n^{\alpha_n}$ be a monomial with $\alpha = (\alpha_1,\cdots,\alpha_n)\in\mathbb{N}^n$  and let $\mathcal{M}_+(\mathcal{X})$ be the set of nonnegative Borel measures supported on $\mathcal{X}$.

Given a nonnegative Borel measure $\xi$ with support on $\mathcal{X}$,
\begin{equation}\label{eq:8}
y_{\alpha} = \int_{\mathcal{X}} x^{\alpha}d\xi
\end{equation}
is the moment of order $\alpha$ of $\xi$.  Let $\mathbf{y} = (y_{\alpha})_{\alpha\in\mathbb{N}^n}$ be the moment sequence of $\xi$ and for a pre-selected positive integer $d$, let $\mathbf{y}_d= (y_{\alpha})_{|\alpha|\leq 2d}$  be the truncated sequence that includes the elements corresponding to $|\alpha|\leq 2d$, where $|\alpha|= \sum_{i=1}^n\alpha_i$. For brevity, we write $\mathbf{y}_d$ simply as $\mathbf{y}$ when the truncated degree is $2d$, and add a special subscript when truncated degree is different. The moment matrix of a polynomial regression model with $n$ factors in $\mathcal{X}$ and highest  degree $d$ is a one to one map to the set
$$\mathbf{C}_{2d}(\mathcal{X}):= \{\mathbf{y}\in\mathbb{R}^q: \exists\xi\in\mathcal{M}_+(\mathcal{X})\ s.t.\ y_{\alpha} = \int_{\mathcal{X}} x^{\alpha}\xi(dx)\},$$
with  $q=C_n^{n+2d}$, which is a set of moment sequences. Here is a simple example. Suppose we have $n=2$ factors with highest degree $d = 1$. The moment sequence is a vector with $C_2^4= 6$ elements and the moment matrix is
$$M(\xi) = \begin{pmatrix}
 y_{(0,0)} & y_{(1,0)} & y_{(0,1)}\\
y_{(1,0)} & y_{(2,0)} &y_{(1,1)}\\
 y_{(0,1)} &y_{(1,1)} & y_{(0,2)}
 \end{pmatrix}.$$
 Notice that $M(\xi)$ is the same as $M$ in section 3.1, where we now denote it as a function of $\xi$ to emphasize its relationship with designs. The matrix is symmetric and is determined by the six elements in the vector $\mathbf{y} = [y_{(0,0)},  y_{(1,0)},
y_{(0,1)},$ $y_{(2,0)},y_{(1,1)}, y_{(0,2)}].$ Therefore, each moment matrix is uniquely defined by a moment sequence, and we denote it as $M_d(\mathbf{y})$, where $d$ is the highest degree of factors.

\subsubsection{Semidefinite approximations of moment cone}
We constrain the design space $\mathcal{X}$ by a set of inequalities constraints to use the hierarchy approximation method. Let $g_i(x)$ be given polynomials of degree $d_i$ and let $$\mathcal{X}:=\{x\in\mathbb{R}^n: g_i(x)\geq 0, i=1,\cdots, L\}.$$
Given $d$, we define a localizing matrix for a given multi-factor polynomial $f(x) = \sum_{|\alpha|\leq 2s}f_{\alpha}x^{\alpha}$ of degree $2s$ by a sequence $\mathbf{y}_s = (y_{\alpha})_{|\alpha|\leq (2d+2s)}$. The entry correspond to $(\alpha,\beta)$ of this localizing matrix $M_d(f\mathbf{y}_s)$ has the form $\sum_{|\gamma|\leq 2s}f_{\gamma}y_{\gamma+\alpha+\beta}.$

By Putinar's theorem \citep{lasserre2009moments}, a moment cone can be approximated by a hierarchy of semidefinite cones. Given $g_i$ as described above,
let $v_i = \lceil d_i/2\rceil, 1\le i\le L$, $q=C_n^{n+2d}, v=C_n^{n+2(d+\tau)}$ and $\tau\in\mathbb{N}$ is a pre-selected relaxation order.  Define
\begin{equation}\label{sdp}
\begin{aligned}[b]
\mathbf{C}_{2(d+\tau)}^{SDP}(\mathcal{X}):=\{\mathbf{y}\in\mathbb{R}^q: \exists \mathbf{y}_{\tau}\in\mathbb{R}^v\ s.t.\ (\mathbf{y}_{\tau})_{|\alpha|\leq 2d} &= \mathbf{y},
 M_{d+\tau}(\mathbf{y}_{\tau})\succeq 0,\\
 & M_{d+\tau-v_i}(g_i\mathbf{y}_{\tau})\succeq 0, \forall i\},
 \end{aligned}
 \end{equation}
where $\mathbf{y}_{\tau} = (y_{\alpha})_{|\alpha|\leq(2d+2\tau)}$  and $(\mathbf{y}_{\tau})_{|\alpha|\leq 2d}$ is a vector composed of the elements in $\mathbf{y}_{\tau}$ corresponding to $|\alpha|\leq 2d$.
Because $\mathbf{C}_{2d}(\mathcal{X})\subseteq \mathbf{C}_{2(d+\tau)}^{SDP}(\mathcal{X})$, this approach is called an outer approximation. By \cite{de2017d},
$\mathbf{C}_{2d}(\mathcal{X})\subseteq \cdots \subseteq \mathbf{C}_{2(d+1)}^{SDP}(\mathcal{X})\subseteq \mathbf{C}_{2d}^{SDP}(\mathcal{X})$
and the hierarchy converges, which means  that $\mathbf{C}_{2d}(\mathcal{X})=\overline{\cap_{\tau=0}^{\infty}\mathbf{C}_{2(d+\tau)}^{SDP}(\mathcal{X})}$. In what is to follow, we now use this fact and develop a semidefinite programming approximation scheme for the moment cone constraint problem.
First rewrite our optimization problem as

\begin{equation}\label{eq:12}
\begin{aligned}
& \underset{\mathbf{y}, t_{jk}, z_{jk}}{\text{maximize}}
& & \min_{j>k} (-t_{jk})
    - \delta_{\Theta_j}^*(z_{jk}) - \delta_{\Theta_k}^*(-z_{jk}) \\
& \text{subject to}
&&
 \left[\begin{array}{cc} M_{d}(\mathbf{y}) & z_{jk} \\ z_{jk}^T & 4t_{jk}
 \end{array}\right] \succeq 0, \quad j>k \\
&&&\mathbf{y}\in \mathbf{C}_{2d}(\mathcal{X}),\quad y_0=1.
\end{aligned}
\end{equation}
By Theorem 4.3 in \cite{de2017d}, constraint (\ref{eq:12}) can be approximated by a series of semidefinite constraints defined as (\ref{sdp}).
When the relaxation order $\tau\rightarrow\infty$, the optimization problem (\ref{eq:12}) is equivalent to

\begin{equation}\label{14}
\begin{aligned}
& \underset{\mathbf{y}, t_{jk}, z_{jk}}{\text{maximize}}
& & \min_{j>k} (-t_{jk})
    - \delta_{\Theta_j}^*(z_{jk}) - \delta_{\Theta_k}^*(-z_{jk}) \\
& \text{subject to}
&&
 \left[\begin{array}{cc} M_{d}(\mathbf{y}) & z_{jk} \\ z_{jk}^T & 4t_{jk}
 \end{array}\right] \succeq 0, \quad j>k \\
&&&\mathbf{y}\in \mathbf{C}_{2(d+\tau)}^{SDP}(\mathcal{X}), \quad y_0=1,
\end{aligned}
\end{equation}
which is a semidefinite programming problem with respect to $\mathbf{y}$.

In practice, it may not be clear whether a certain relaxation has achieved convergence or not. This is especially so  in high-dimensional problems, where it may be hard to discern whether the current relaxation is enough  or a still higher relaxation is needed.  In the latter case, greater computational effort is required to handle the additional variables and the larger moment matrix. When the criterion is convex,  we resort to  an equivalence theorem  based on the duality theorem, to check whether a design is globally optimal or not. Subsection 3.4 provides details.

\subsection{Solution extraction}
The solution extraction problem is an $\mathcal{A}$-truncated $X$-Moment problem studied by \cite{nie2014mathcal}. It concerns whether a given vector $\mathbf{y}_{\mathcal{A}}$ admits an atomic measure $\mu$ in $\mathcal{X}$. The problem can be proposed as one of finding an atomic measure $\mu$ that satisfies the constraints:
$$\int_{\mathcal{X}} x^{\alpha}\mu (dx) = y_\alpha\quad \forall \alpha\in\mathcal{A},$$
where $\mathcal{A}$ is a finite set indicating the power set of the vector $\mathbf{y}_{\mathcal{A}}=(y_{\alpha})_{\alpha\in\mathcal{A}} \subset \mathbb{N}^n$.

From \cite{nie2014mathcal} and \cite{de2017d}, the optimal design $\xi$ can be obtained by solving a hierarchy of $\mathcal{A}$-truncated $X$-Moment problems given by
\begin{equation}\label{eq:17}
\begin{aligned}
& \underset{\mathbf{y}_r\in\mathbb{R}^{C^{n+2(d+r)}_n}}{\text{minimize}}
& & \text{trace}(M_{d+r}(\mathbf{y}_r))\\
& \text{subject to}
&&M_{d+\tau}(\mathbf{y}_r)\succeq 0\\
&&&M_{d+\tau-v_i}(g_i\mathbf{y}_r)\succeq 0,\ i=1,\cdots L\\
&&&(\mathbf{y}_r)_{|\alpha|\leq 2d} = \mathbf{y}^*
\end{aligned}
\end{equation}
where $v_i=\lceil d_i/2\rceil$,
 $\mathbf{y}^*$ is the optimal value from (\ref{14}) and $r$ is another user-selected relaxation order larger than $\tau$.   We then increase the value of $r$ by one each time until a solution $\mathbf{y}_r^*$ of (\ref{eq:17}) is found and it satisfies the rank condition
$$\text{rank}\ M_{d+r}(\mathbf{y}_r^*) = \text{rank}\ M_{d+r-v}(\mathbf{y}_r^*)$$
where $v = \max_i v_i$. \cite{nie2014mathcal} proved that when $\mathbf{y}^*$ is a moment vector, the rank condition will definitely be satisfied when $r$ is large enough. After  $\mathbf{y}_r^*$ is obtained, we apply the methods in \cite{henrion2005detecting} to extract the support points of the optimal design. We then calculate the weights using $\mathbf{y}^*$ and support points based on (\ref{eq:8}).

\subsection{Equivalence theorem}\label{equiv}
When the objective function to optimize is a convex function, an equivalence theorem may be used to check whether a given design is a global optimum. Each convex functional has a unique equivalence theorem which is based on the sensitivity function of the design under investigation.  The sensitivity function of the design is the directional derivative of the convex functional in the direction of a degenerate design at the point $x$ and evaluated at the design.  The equivalence theorem states that if the design is optimal, its sensitivity function is non-positive throughout the design space with equality at its support points; otherwise, the design is not optimal among all designs on the given design space.  When the design space is an interval with one factor, the sensitivity function is univariate and can be easily plotted on the design space for a visual appreciation.  
For example, when we have one factor and the vector of regression functions $f(x)$ has $l$ linearly independent components with homoscedastic errors, the equivalence theorem for $D$-optimality states the design $\xi_D$  is $D$-optimal among all designs on $\mathcal{X}$ if and only if $f^T(x)M(\xi_D)^{-1}f(x)-l\le 0$ for all $x\in \mathcal{X}$. In this case, the equivalence theorem is closely related to \textsl{Christoffel polynomials}, see \cite{hess2017some}.

To discriminate between two models using T-optimality, a direct calculation shows the criterion is proportional to the non-centrality parameter defined in subsection 3.1, which is
$$\Phi(\xi) = \min_{\theta_{12}} \theta_{12}^TM(\xi)\theta_{12}={\theta_{12}^*}^TM(\xi)\theta_{12}^*$$
where $\theta_{12} = \theta_1 - \theta_2$ and $M(\xi) = \int_{\mathcal{X}} f(x)f(x)^Td\xi$. Manifestly, the function $\Phi(\xi)$ is a concave function of $\xi$ over the set of all designs on $\mathcal{X}$.  If $\xi^*$ is the optimal discrimination design and  $\xi'$ is the degenerate design at the point $x$  near the optimum $\xi^*$, we have
$$\frac{\partial}{\partial\alpha}\{\theta_{12}^{*T}M[(1-\alpha)\xi^* + \alpha\xi']\theta_{12}^*\}|_{\alpha = 0}\leq 0.$$
This implies that
$$\theta_{12}^{*T}M(\xi')\theta_{12}^*-\Delta(\xi^*)\leq0,$$
where $\Delta(\xi^*) = \theta_{12}^{*T}M(\xi^*)\theta_{12}^*$  is the optimal value. If we let $\phi(x, \xi^*) = \{\eta_1(x, \theta_1^*) - \eta_2(x, \theta_2^*)\}^2$, the equivalence theorem states that  $\xi^*$ is an optimal discrimination design if and only if
\begin{equation}\label{eq:16}
\phi(x, \xi^*)-\Delta(\xi^*)\leq0~~\text{for $x\in \mathcal{X}$},
\end{equation}
with equality at the support points of the T-optimal design.

For a design with several competing models, the equivalence theorem is a straightforward generalization of that for discriminating between two models \citep{atkinson1975design}. If there is only one closest distance between different combinations of models and this occurs, say, between the null model and model $k$, then the equivalence theorem becomes
$$\phi_{k}(x, \xi^*)-\Delta(\xi^*)\leq0 $$
where $\phi_{k}(x, \xi^*) = \{\eta_1(x) - \eta_k(x, \theta_k^*)\}^2$ with  equality at the support points of the optimal design. For design problems where there are multiple points in the parameter space that achieve the smallest distance, see  the more complicated equivalence theorem in Theorem 1 of \cite{atkinson1975b}.

We note that the design obtained by solving (\ref{14}) and (\ref{eq:17}) with large enough relaxation orders $\tau$ and $r$ is a T-optimal design. The sensitivity function of the design provides us with an additional tool to confirm its optimality via the equivalence theorem.  Since we only require the optimal moment matrix $M(\xi^*)$ to obtain both $\Delta(\xi^*)$ and $\theta_{jk}^*$  to use the equivalence theorem, we can easily obtain information on whether the relaxation order $\tau$ is big enough for the semidefinite relaxation mentioned in (\ref{14}) to apply.



\section{Algorithm}
Our algorithm requires that the optimal discrimination design problem is properly formulated as a convex optimization problem.   The proposed algorithm has
three major steps: i). Solve the relaxed SDP problem in (\ref{14}); ii). Extract the optimal design from the optimal moment matrix obtained from step 1 and iii). Use the equivalence theorem to validate the optimality of the design. 
These steps are described below, along with a set of pseudo codes for the whole procedure in the summary table labeled 4. We remind the reader that Step 3 is desirable but not a necessary step.
\begin{step}
\caption{\textit{RelaxedSDP}}
\begin{algorithmic} 
\REQUIRE Relaxation order $\tau\geq 0,\{\Theta_k,k=1,\cdots,K\},\{g_i(x),i=1\cdots,L\}$, \\basis function $f(x)$
\STATE Solve (\ref{14}) using Gloptipoly3 package and SDP solvers. 
\ENSURE Optimal matrix $M^*(\xi)$ and optimal value $t^*$. 
\end{algorithmic}
\end{step}
\begin{step}
\caption{\textit{Extraction}}
\begin{algorithmic} 
\REQUIRE Relaxation order $r> 0, M^*(\xi), \{g_i(x),i=1\cdots,L\}$, basis function $f(x)$
\STATE Solve (\ref{eq:17}) using Gloptipoly3 package and SDP solvers. 
\ENSURE Support points of optimal design $(x_1,\cdots x_m)$
\end{algorithmic}
\end{step}
\begin{step}
\caption{\textit{EquivalenceTheorem}}
\begin{algorithmic} 
\REQUIRE $M^*(\xi), \{\Theta_k,k=1,\cdots,K\},\{g_i(x),i=1\cdots,L\}$
\STATE Check if (\ref{eq:16}) holds for every single point on design space $\mathcal{X}$.
\ENSURE True (equivalence theorem is not violated) or False
\end{algorithmic}
\end{step}

\begin{summary}
\caption{\textit{T-optimal Design using SDP relaxation}}
\begin{algorithmic} 
\REQUIRE $\{\Theta_k,k=1,\cdots,K\},\{g_i(x),i=1\cdots,L\}$, basis function $f(x)$
\STATE $t^*_{-1}=\infty$, $\tau = 0$
\WHILE{True}
\STATE $M^*(\xi), t_{\tau}^*$ = \textit{RelaxedSDP}($\tau, \Theta_k,g_i(x),f(x)$)
\IF{$t^*_{\tau} == t^*_{\tau-1}$} \STATE Break
\ELSE \STATE $\tau = \tau + 1$
\ENDIF
\ENDWHILE
\STATE $r = \tau+1$
\WHILE{True}
\IF{\textit{Extraction($r, M^*(\xi), \{g_i(x),i=1,\cdots,L\}, f(x))$} succeed} \STATE $\xi^* =$\textit{Extraction}($r, M^*(\xi), \{g_i(x),i=1,\cdots,m\}, f(x))$ \STATE Break
\ELSE \STATE $r = r+1$
\ENDIF
\ENDWHILE
\STATE Validation = \textit{EquivalenceTheorem}($M^*(\xi),\{\Theta_k,k=1,\cdots,K\},\{g_i(x),i=1\cdots,L\}$)
\IF{Validation}\STATE Return optimal design
\ENDIF
\ENSURE Optimal design $\xi^*$, optimal value $t^*$
\end{algorithmic}
\end{summary}

\newpage

\section{Examples}
\label{sec:exam}
We provide several examples to demonstrate our proposed methodology.  Some of the examples are selected to show our algorithm provides the same T-optimal designs reported in the literature and others to show our methodology is more flexible or efficient than current methods. 

Convex optimization has been studied for decades and has many modeling tools and solvers to solve convex optimization problems efficiently. There are two classes of optimization softwares, one class serves as modeling tools and the other serves as solvers. Several modeling tools for convex problems are available for academic use and they include CVX \citep{grant2008cvx}, YALMIP \citep{lofberg2004yalmip}, CVXPY \citep{diamond2016cvxpy}, LMI Lab \citep{gahinet1994lmi}, ROME \citep{goh2011robust} and AIMMS \citep{bisschop2006aimms}.  Both CVX and YALMIP can directly call popular solvers for SDP and they include Mosek \citep{mosek}, SeDuMi \citep{sturm1999using}, SDPT3 \citep{toh1999sdpt3}, etc.


\subsection{Numerical Examples}
We now apply the moment relaxation method to solve T-optimal design problems for a variety of situations. All of the examples are modeled by GloptiPoly3 \citep{henrion2009gloptipoly} and YALMIP, and solved by MOSEK 7 or SeDuMi 1.3 in the MATLAB 2014a environment.
For each example, we list results in Table \ref{tab:outcomes}, including the dimension of the design space, i.e. the number of factors in the study, the optimal value of the design criterion, CPU time for solving the problem, number of unknown parameters in the problem, number of support points in the optimal discrimination design, the optimal design and the relaxation orders of $\tau$ and $r$. All results were obtained using a 16 GB RAM Intel Core i7 machine running 64
bits Mac OS operating system with 2.5 GHz.

We present six examples with various setups to illustrate flexibility of our approach and its advantages over current methods. Except for one case,  all factors are restricted to the design interval $[-1,1]$, in this case, the polynomial inequality constraints would be $g(x_i) = 1-x_i^2, \forall i = 1, \cdots, n$.  Our examples include incomplete polynomial models, different numbers of factors in the models up to 7 and discrimination among two or three polynomial models. Some examples do not require the null model be completely specified and allows for different levels of uncertainty for each unknown coefficient in the polynomial model.  

The first two univariate examples are selected from \cite{duarte2015semi}, from which we observe that our algorithm is more computationally efficient.  Example 3 concerns an optimal design discrimination problem  for two models, where possible values for the coefficients in each model  are confined to a user-specified region. Example 4 and Example 5 show different solvers may be more appropriate for different situations. Example 6 is an optimal design discrimination problem for three competing models. Notice that  we have listed models as $\eta_1$, $\eta_2$ and $\eta_3$ but this numbering is arbitrary as  there is little difference whether one model is labelled first or not.

\bigskip

\textbf{Example 1:}  This problem has one factor and the two mean functions have degrees up to 2.

$\eta_1 = 1 + x + x^2$

$\eta_2 = \theta_{20} + \theta_{21}x$

$\mathcal{X} = [-1, 1], \Theta_2 \in [0,4]^2.$

\bigskip

A direct application of our algorithm produced the same optimal design reported as Case (2) in Duarte et al. (2015) which was found by semi-infinite programming method.   Their CPU time is much longer than the CPU time required for our algorithm to generate the design, which is displayed in the first row of Table 1.  Both algorithms gave the same optimal value.

\bigskip

\textbf{Example 2:} This problem has one factor and the two mean functions have degrees up to 5.

$\eta_1  = 1 + x + x^2 + x^3 + x^5$

$\eta_2 = \theta_{20} + \theta_{21}x+\theta_{22}x^2+\theta_{23}x^3$

$\mathcal{X} = [-1, 1], \Theta_2 \in [0,4]^4.$

\bigskip

This example is Case (4) in \cite{duarte2015semi} where they solved the problem using a semi-infinite programming approach by first searching among designs with a predetermined number of points.  This number is usually taken to be equal to the number of parameters in the model plus $1$, which is $5$ in this case. If the algorithm does not converge, the search expands to among designs supported at one more point. It is common that only a copy of such expansions is needed.
For this example,
they found two asymmetric supported at $5$ points  optimal designs
and their sensitivity functions show they satisfy the equivalence theorem.
Interestingly, our algorithm produced a $6$-point symmetric design shown in the second row of Table 1.  It is, numerically, a convex combination of the two asymmetric designs and it has the same optimality value as the ones found by \cite{duarte2015semi}. A major difference is that our algorithm produced the optimal design much more efficiently in that it only required about 1.1 sec of CPU time compared to their 335 seconds of CPU time required by the semi-infinite programming method. However, our algorithm only finds an optimal design and does not find another when the optimal designs are not unique.
 We note that $\eta_1$ has the missing term $x^4$ in its mean function and so it is an example of an incomplete polynomial model as discussed in \cite{dette2012t}.
\bigskip

\textbf{Example 3:} This problem has two factors and the two mean functions have degrees up to 2.  Both null and alternative model have unknown coefficients in the polynomial models.

$\eta_1 = \theta_{10} + \theta_{11}x_1 + \theta_{12}x_2 + \theta_{13}x_1^2$

$\eta_2 = \theta_{20} + \theta_{21}x_1 +\theta_{22}x_2 + \theta_{23} x_1^2 + \theta_{24} x_2^2 + \theta_{25} x_1x_2$

$\mathcal{X} = [0, 4]\times [-1,1], \Theta_1\in[0,4]^4,\ \Theta_2 \in [0,2]^4\times [1,4]^2.$

\bigskip

Unlike the first two examples,   the null model in this example is not fully specified. The parameter spaces of $\theta_{24}$ and $\theta_{25}$ do not include $0$ so that the two models are not identical; otherwise, the two models are indistinguishable. This example also shows the flexibility of our methodology.  One of the design spaces is no longer $[-1, 1]$ and the parameter spaces for the various model parameters are different, and neither of the models is fully specified. The left panel in Figure 1 shows the sensitivity function of the design found by GloptiPoly3. The optimal discrimination design and some characteristics of the  design setup are reported in the 3rd row of Table 1. 
\bigskip

\textbf{Example 4:}  The problem has 3 factors and the null model is fully specified with some two factor interactions and  the three factor interaction term.

$\eta_1 = 1 + x_1 + x_2 + x_3 + x_1 ^ 2 + x_2^2 + x_3^2 + x_1x_2^2 + x_1x_2x_3$

$\eta_2 = \theta_{20} + \theta_{21}x_1 +\theta_{22} x_2 + \theta_{23}x_3 + \theta_{24}x_1 ^ 2 + \theta_{25}x_2^2 + \theta_{26}x_3^2$

$\mathcal{X} = [-1, 1]^3, \Theta_2 \in [0,4]^7.$

\bigskip

We solve this optimal discrimination design using \textsl{SeDuMi}.  \textsl{SeDuMi} provides the solution quickly and precisely.  
When models have multiple factors, it is not easy to display and appreciate the properties of the sensitivity function of a design over the whole high dimensional design space.  One option  is to discretize the design space into a set of fine grid points and number them. 
We then order each index on the horizontal and plot the sensitivity function versus the ordered indices to confirm whether the number of peaks is equal to the number of support points in the generated design and the peaks correspond to the indices that match the support points of the design.  
We employ such a strategy to check optimality in this and other examples with multiple factors.
The plot is shown on right panel in Figure 1. Another option is to order the points in a factorial order used in \cite{fedorov2013optimal} for a more specialized setup.

\bigskip
\begin{figure}
\begin{tabular}{cc}
\includegraphics[width=0.5\linewidth]{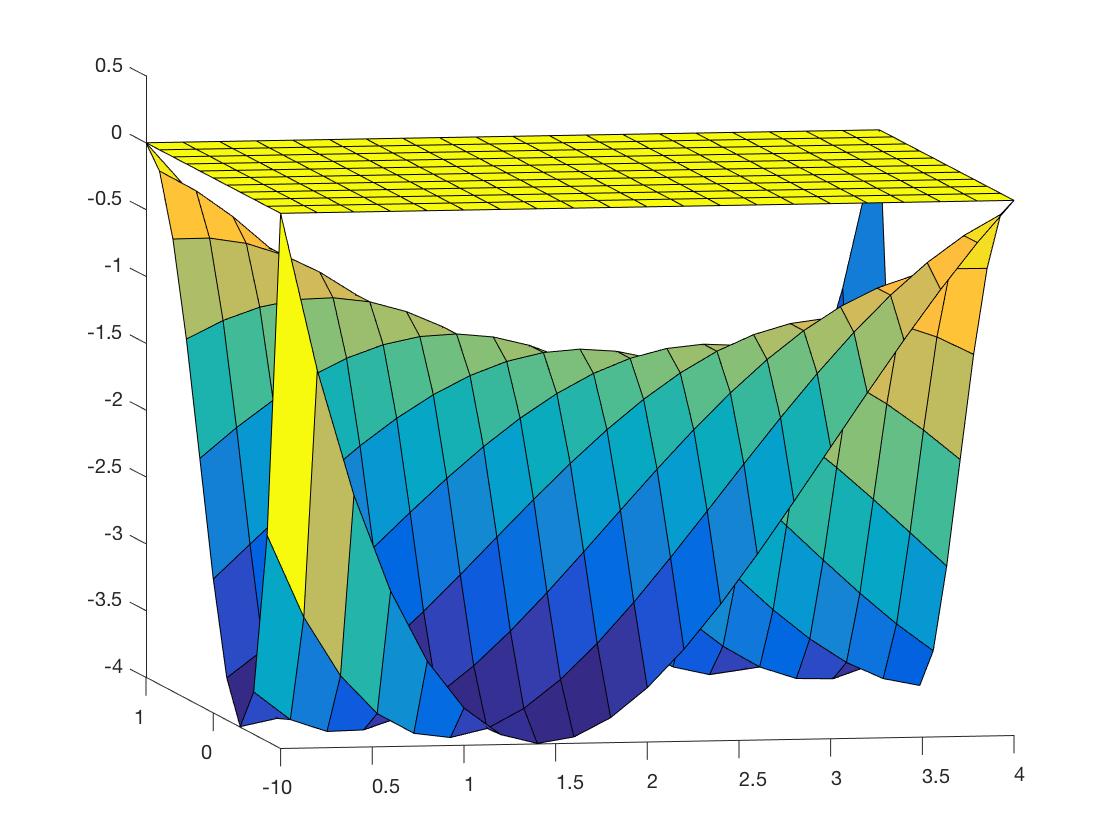} &
\includegraphics[width=0.5\linewidth]{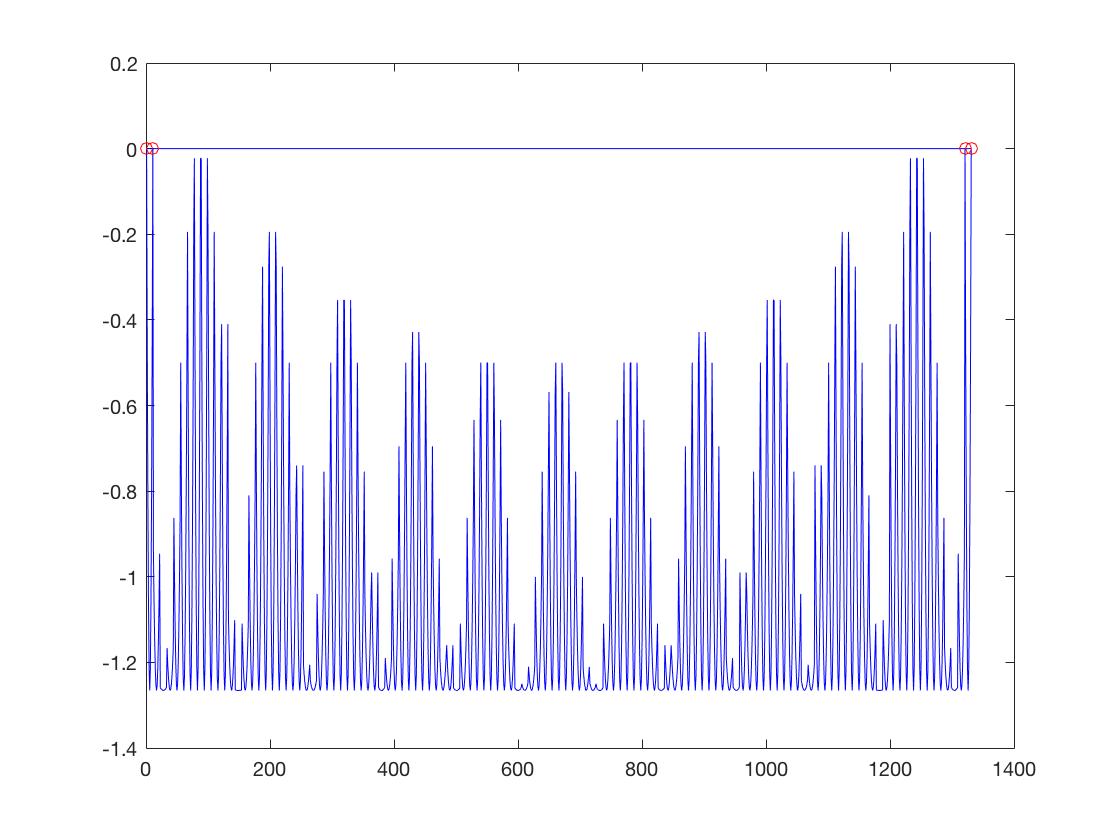}
\end{tabular}
\caption{Plots of the sensitivity functions of the two designs found by our algorithm for Example 3 with two factors (left) and Example 4 with three factors (right). Both plots confirm the optimality of the designs.}
\end{figure}










\textbf{Example 5:} This example has seven factors.  The null model is fully specified with complete first and second order  terms and all pairwise interaction terms.  The alternative model has less factors and no interaction terms.




$\eta_1 = 1 + x_1 + \cdots + x_7 + x_1 ^ 2 + x_1x_2+ \cdots + x_6x_{7} + x_{7}^2$

$\eta_2 = \theta_{20} + \theta_{21}x_1 +\theta_{22} x_2 + \theta_{23}x_3 + \theta_{24}x_4+ \theta_{25}x_5 +  \theta_{26}x_6 +\theta_{27}x_1 ^ 2+\theta_{28}x_2 ^ 2+ $\\
$\indent\quad\quad \theta_{29}x_3 ^2+\theta_{2,10}x_4^2$

$\mathcal{X} = [-1, 1]^7, \Theta_2 \in [0,4]^{11}.$

\bigskip

In this example, the smallest rank relaxation does not work and we have to resort to using higher order relaxation, including SDP relaxation order $\tau = 1$ and extraction relaxation order $r = 2$.
 \textsl{Mosek}  only takes several minutes  to complete the extraction process and produce a solution. 
We observe that when the mean response is not symmetrically represented by the factors in the polynomial model, the distribution of the support points of the resulting optimal discrimination design is also  asymmetric.

\bigskip



\newpage 

\textbf{Example 6:} This example has three models for discrimination.  The null model is fully specified and has 3 factors up to order 2 with all pairwise interaction terms.  The other 2 alternative models are  additive; one has only first order terms and the other has up to second order terms.

$\eta_1 = 1 + x_1 +   x_2 + x_3 + x_1^2 + x_1x_2 + x_1x_3  + x_2^2+ x_2x_3+x_3^2$

$\eta_2 = \theta_{20} + \theta_{21}x_1 +\theta_{22} x_2 + \theta_{23}x_3$

$\eta_3 = \theta_{30} + \theta_{31}x_1 +\theta_{32} x_2 + \theta_{33}x_3 + \theta_{34}x_1^2 +\theta_{35} x_2^2 + \theta_{36}x_3^2$

$\mathcal{X} = [-1, 1]^3, \Theta_2 \in [1,2]^{4}, \Theta_3 \in [1,2]^7.$

\bigskip

From this example, we also observe that  our algorithm can not only solve the classical problem where the `true model' has known parameters and the `test model' has unknown parameters, but it can also discriminate two models with uncertain parameters. Specifically, this  example with three possible models can be solved by introducing an auxiliary variable as the lower bound for three pairwise distances, and maximize over that auxiliary variable 
because we aim to maximize the minimal squared distance. \\
\indent In this example, the optimal design has equal distances between $\eta_1$, $\eta_3$ and between $\eta_2$ and $\eta_3$. More specifically, under the optimal design, the distance between $\eta_1$ and $\eta_3$ is $4$, while that between $\eta_1$ and $\eta_2$ is $13$, and between $\eta_2$ and $\eta_3$ is $4$. Noticing that the optimal $\theta_{34}^*=\theta_{35}^*=\theta_{36}^*=1 $ when comparing models $2$ and $3$, this makes sense since between $\eta_1$ and $\eta_3$ their differences are in the second order interactions and the differences between $\eta_2$ and $\eta_3$ is only in the second order terms of the factors. Since there are no differences among those factors, it is reasonable that their interactions and their own second order terms have the same effects.\\
\indent Figure 2 shows the directional derivative of the criterion evaluated at the optimal design. To show that there are exactly $8$ design points in each plot, we discretize the space fairly sparse so that the design points  can be observed clearly on the plots. In this case, we only sample $5$ points uniformly from $x_1,x_2$ and $x_3$ to make the plot clear. However, when we ascertain optimality of a design using the equivalence theorem in practice, we need to discretize the space as dense as possible to verify that the sensitivity function has the same peak values at its support points. There are two plots because there are two competing pairs of models in this example with the same minimal distance.  Accordingly, the equivalence theorem requires that we require two sensitivity plots to confirm optimality of the optimal discrimination design.  Figure 2 confirms that  the generated design shown in Table 1 line 6  satisfies the equivalence theorem and so the design is optimal for discriminating among the 3 models in the problem.

This example can also be converted to a problem in format (\ref{eq:3}) by choosing weights for the comparisons between different models. In \cite{atkinson1975b}, the concept of different weights for each pair was proposed, but no specific examples were provided. As an illustration, if we assign the weights to be $w_{12}=0.2,w_{13}=0.2,w_{23}=0.6$, in this example we obtain from our algorithm an equally weighted optimal design at two end points $(1,1,1)$ and $(-1,-1,-1)$, and the optimality value is $7.4$.


\begin{table}
\centering
{\footnotesize
\caption{Features of the Discrimination Design Problems and Optimal Designs for Examples 1-6 in the order listed below} \label{tab:outcomes}
\begin{tabular}{ |p{1.2cm}||p{1.5cm}|p{1.4cm}|p{1.4cm}|p{5.5cm}|p{1.5cm}|p{1.5cm}|}
 \hline
  Number&  Number of &CPU & Number of &$T$-optimal discrimination & Optimal &Relaxation \\
 of factors &parameters& time (sec)& support points&design & criterion ~~~value &order pairs ($\tau,r$)\\
 \hline
1 & 2 & 0.89  & 3 & $\begin{pmatrix}
  -1 & 0 & 1 \\
  0.25 & 0.5 & 0.25
 \end{pmatrix}$ & 0.25 &(0,1)\\
 \hline
  1 & 4 & 1.11  & 6 & $\begin{pmatrix}
  -1 & -0.809 & -0.309 & 0.309 & 0.809 & 1 \\
  0.099 & 0.198 & 0.199 & 0.200 & 0.202 & 0.101
 \end{pmatrix}$& 0.003906&(0,1)\\
\hline
 2 & 4,6 & 0.85  & 4 & $\begin{pmatrix}
  0 & 0 & 4 & 4  \\
  -1 & 1 & -1 & 1 \\
  0.25 & 0.25 & 0.25 & 0.25
 \end{pmatrix}$ & 4&(0,1)\\
\hline
 3 & 7 & 1.02  & 8 & $\begin{pmatrix}
  -1 & 1 & -1 & 1  \\
  -1 & -1 & 0.5 & 0.5 \\
    -1 & -1 & -1 & -1 \\
  0.1253 & 0.1253 & 0.1251 & 0.1251   \\
  \hline
 -1& 1 & -1 & 1\\
  -0.5&  -0.5 & 1 & 1\\
   1&1&1&1\\
   0.1249& 0.1249&0.1247&0.1247
 \end{pmatrix}$ & 1.26562 & (0,1)\\
 \hline
  7 & 11 & 734  & 12 & See $\xi^* below$ & 175.5728&(1,2)\\
\hline
3 & 4,7 & 3.98  & 8 & $\begin{pmatrix}
  -1 & -1 &1 & -1  \\
  -1 & -1 &-1 & 1  \\
  -1 &  1 &-1 & -1  \\
0.2779&0.0555&0.0554&0.0554\\
\hline
1 & -1 & 1 & 1\\
-1 & 1 & 1 & 1\\
1 & 1 & -1 & 1\\
0.1112&0.1112&0.1112&0.2221
\end{pmatrix}$ & 4 &(0,2)\\
\hline
\end{tabular}}
\end{table}

The optimal design for Example 5 is 
$$
\xi^* = \begin{pmatrix}
-1.0000  &  0.3541  &  0.3541  &  0.3541 &  -0.2639  & -0.2639    \\
   -1.0000 &  -1.0000 &   1.0000 &  -1.0000   & 1.0000 &  -1.0000   \\
   -1.0000 &   1.0000 &   1.0000 & 1.0000 &   1.0000  &  1.0000   \\
   -1.0000 &  -1.0000 &  -1.0000 &   1.0000 &  -1.0000  &  1.0000   \\
   -1.0000 &   1.0000 &  -1.0000 &  -1.0000  &  1.0000  &  1.0000   \\
   -1.0000 &   0.3541  &  0.3541 & 0.3541  & -0.2639 &  -0.2639   \\
   -1.0000 &  -0.8541 &  -0.8541 &  -0.8541 &  -1.0000 &  -1.0000   \\
   \hline
   0.2104 & 0.0569 & 0.0570 & 0.0570 & 0.0396 & 0.0396\\
   \hline
      0.3541  &  0.3541 &  -0.2639 & 0.3541 &  -0.2639 &   1.0000\\
     1.0000 &  -1.0000 &   1.0000 & 1.0000 &   1.0000 &   1.0000\\
    -1.0000 &  -1.0000  &  1.0000 & -1.0000 &  -1.0000  &  1.0000\\
   -1.0000 &   1.0000  &  1.0000 & 1.0000  &  1.0000  & 1.0000\\
     1.0000 &   1.0000 &  -1.0000&  -1.0000  &  1.0000 &   1.0000\\
     0.3541  &  0.3541 &  -0.2639  &  0.3541 &  -0.2639 &   1.0000\\
     -0.8541 &  -0.8541 &  -1.0000  &  -0.8541 &  -1.0000  &  1.0000\\
     \hline
     0.0569 &0.0570 & 0.0396 & 0.0569 & 0.0396 & 0.2896\\
     \hline
 \end{pmatrix}
$$

\newpage

\begin{figure}
\begin{tabular}{cc}
\includegraphics[width=0.5\linewidth]{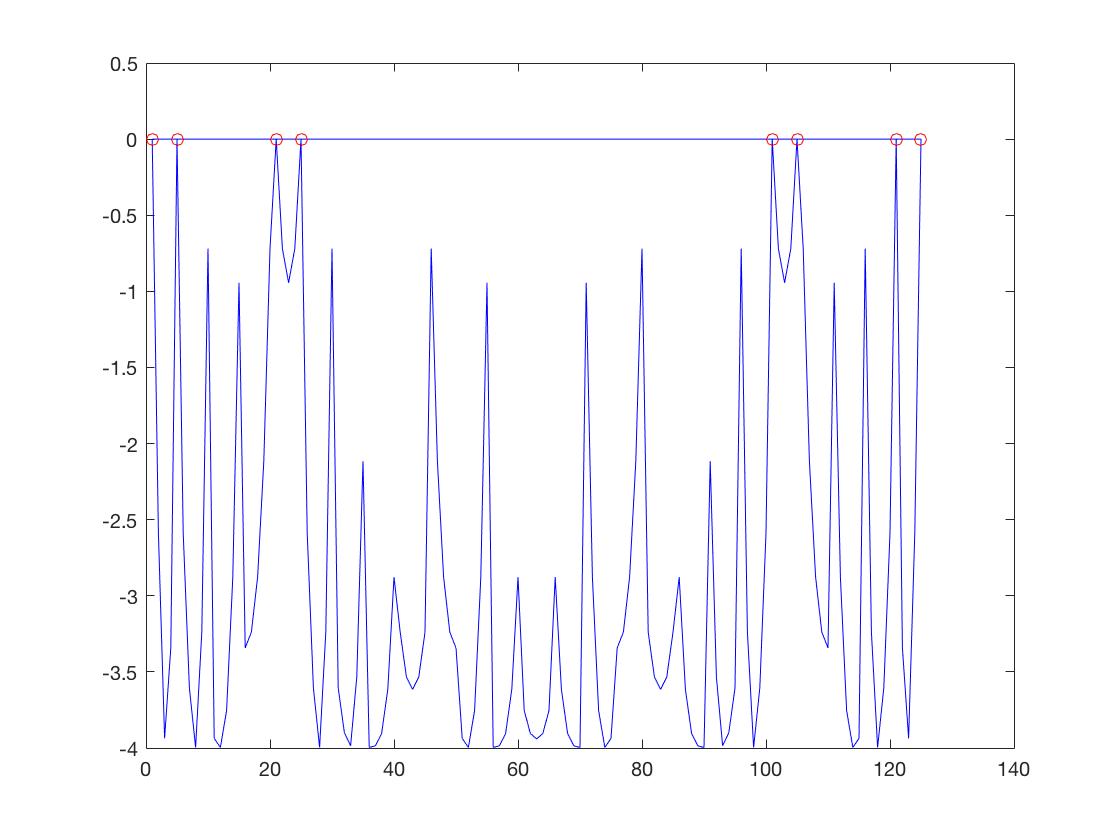} &
\includegraphics[width=0.5\linewidth]{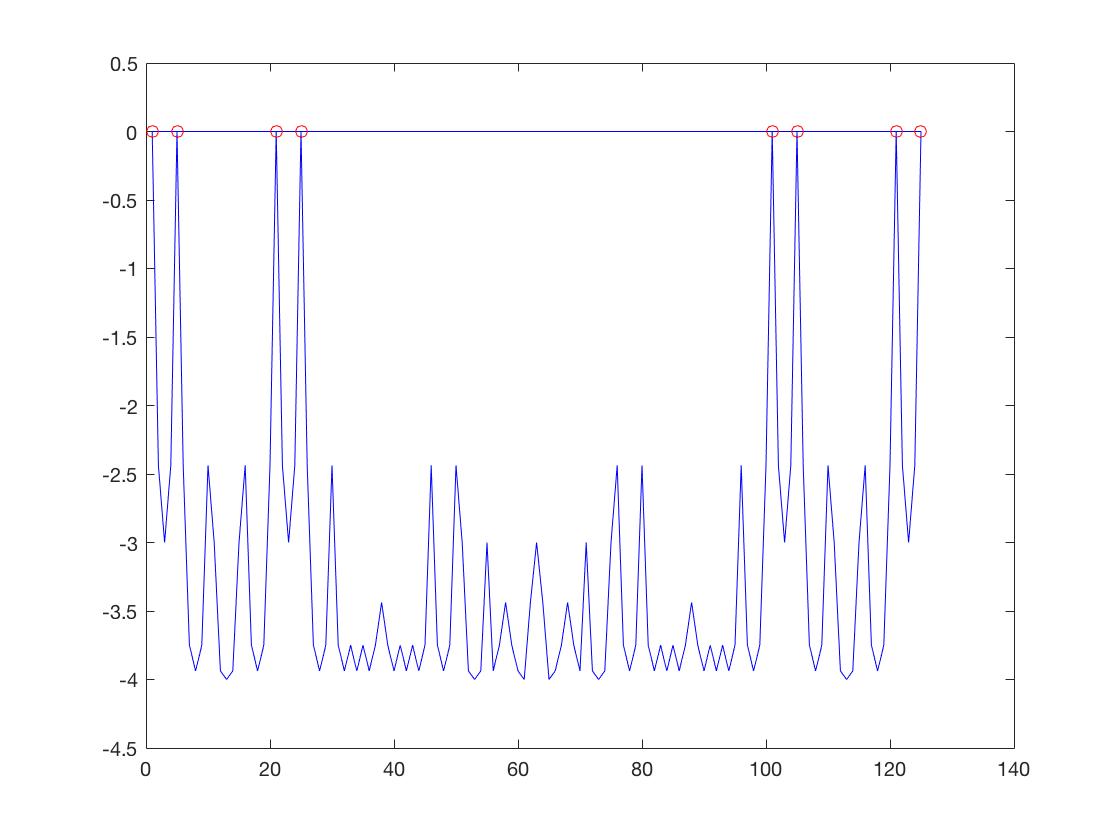}
\end{tabular}
\caption{Plots of sensitivity functions of designs found by our algorithm for Example 6; the left panel is that from the optimal design for discriminating between model 1 and model 3 and the right panel is for discriminating between model 2 and model 3. Both are plotted against a set of discretized points from the design space and they confirm the optimality of the generated designs for this 3-factor experiment with 3 possible mean functions.}
\end{figure}

\section{Discussion}
In this paper, we  generalize the widely used T-optimality criterion for discriminating between 2 or more multi-factor polynomial models  and find an optimal discrimination design by the semidefinite relaxation method. Our experiments use a state-of-the art software \textsl{Gloptipoly3} \citep{henrion2009gloptipoly}  and our experience is that the software typically runs up to a hundred times faster than existing algorithms  when there are two models  to discriminate; for example, see the CPU times in \cite{duarte2015semi}.

Our algorithm  is also more general in that  (i) it is applicable to discriminating three or more models defined on a multi-dimensional  design space with polynomial constraints, (ii) it allows the coefficient in each of the polynomial mean functions has different range spaces, (iii) the null model needs not be completely specified which means that the labeling of the models does not affect the result, (iv) it does not require the design space to be discretized and (v) the user does not have to specify the number of support points of the optimal design in advance.
We also  provide guidelines for the user to determine whether the relaxation order is sufficiently.

We emphasize that optimal designs sought here are very  difficult to study analytically and they have interesting properties.   We provide two illustrations with claims on optimality that we have verified using the equivalence theorems.  First, suppose  we change  the  uncertainty regions for the model parameters   in Example 5 from $[0,4]^{11}$ to $[0,2]^{11}$.  The  generated  $T$-optimal design becomes equally supported at two points at $(1,1,1,1,1,1,1)$ and $(-1,-1,-1,-1,-1,-1,$\\
\noindent$-1)$.    Second, if we change $\eta_2$ in Example 5 to
$$\eta_2 = \theta_{20} + \theta_{21}x_1 ^ 2 +\theta_{22}x_2 ^ 2 +\theta_{23}x_3 ^2+\theta_{24}x_4^2+ \theta_{25}x_5 ^ 2+\theta_{26}x_6 ^ 2+\theta_{27}x_7 ^ 2,~~~~~~ \Theta_2\in [0,4]^8,$$
 the resulting $T$-optimal design has a weight of $0.5$  at the point $(1,1,1,1,1,1,1)$, and the rest of the weights is equally supported at $35$ points.  The point $(-1,-1,-1,-1,-1,-1,-1)$ which was a support point for the original problem is no longer a support point.  Figure 3 shows the locations and the weight distribution of the $T$-optimal discrimination design for this case with the modified $\eta_2$ .  In either of these cases, we were unable to provide an intuitive explanation for the unexpected change in the structure of the $T$-optimal design.
Figure 4 displays the sensitivity function of the design for our second modified example which has $36$ support points. The left panel uses a dense grid to show there is no violation in general, and the right panel uses a sparse grid to show more clearly that there are $36$ support points in total.

\begin{figure}[htp]
\begin{tabular}{c}
\includegraphics[width=0.9\linewidth]{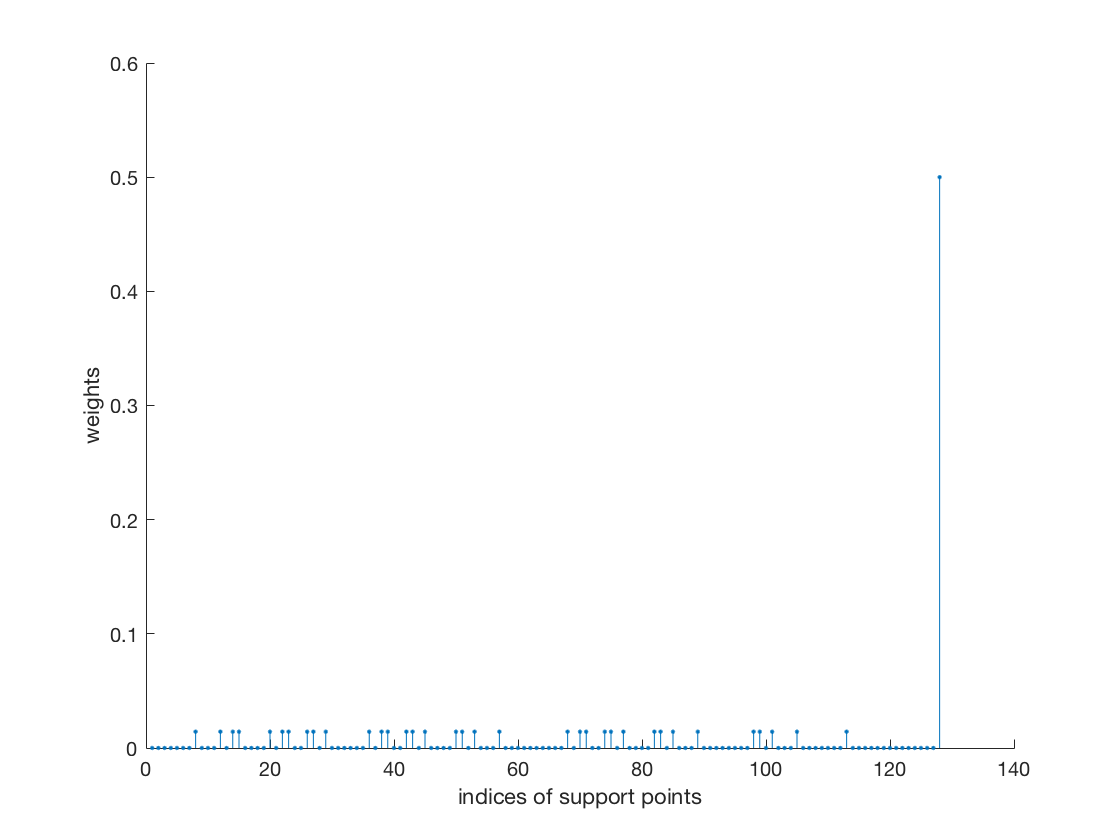}
\end{tabular}
\caption{The $T$-optimal discrimination design found from our algorithm for the modified Example 5.  It has half of its weight at $(1,1,1,1,1,1,1)$ and the rest equally distributed among the 36 indices of the support points shown above in the 7-dimensional design space. }
\end{figure}

\begin{figure}[htp]
\begin{tabular}{cc}
\includegraphics[width=0.5\linewidth]{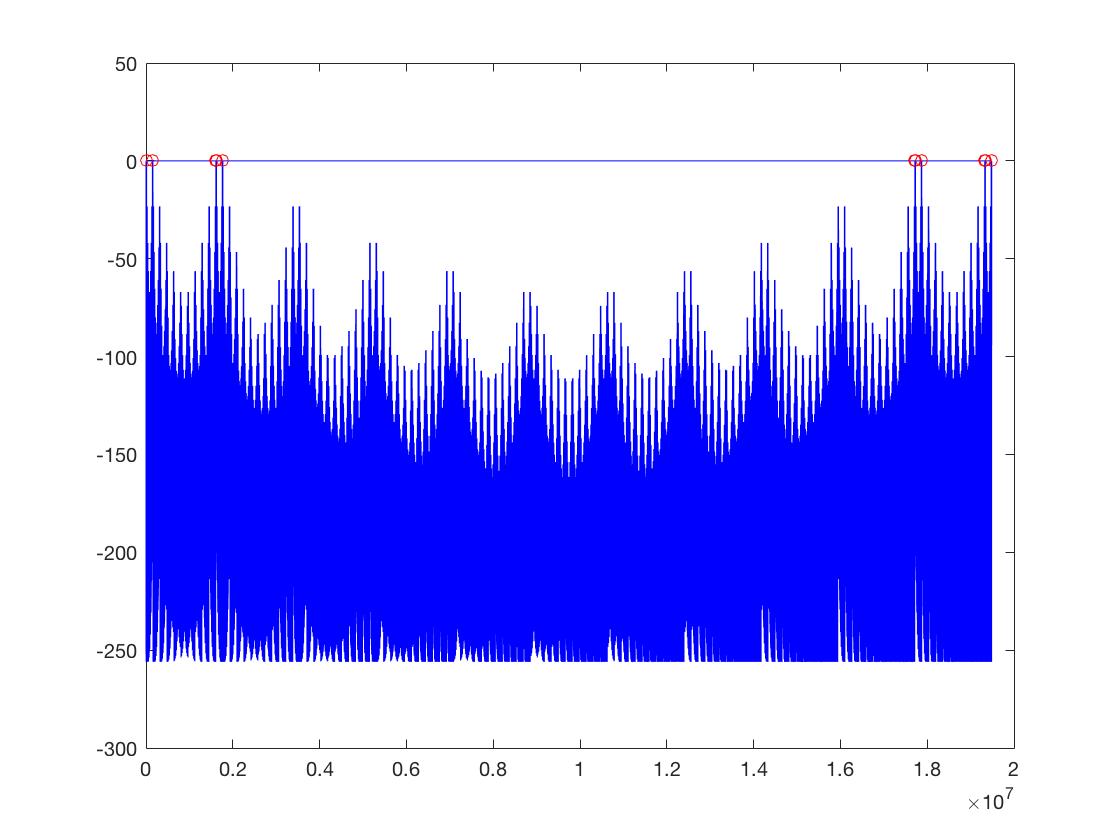}&
\includegraphics[width=0.5\linewidth]{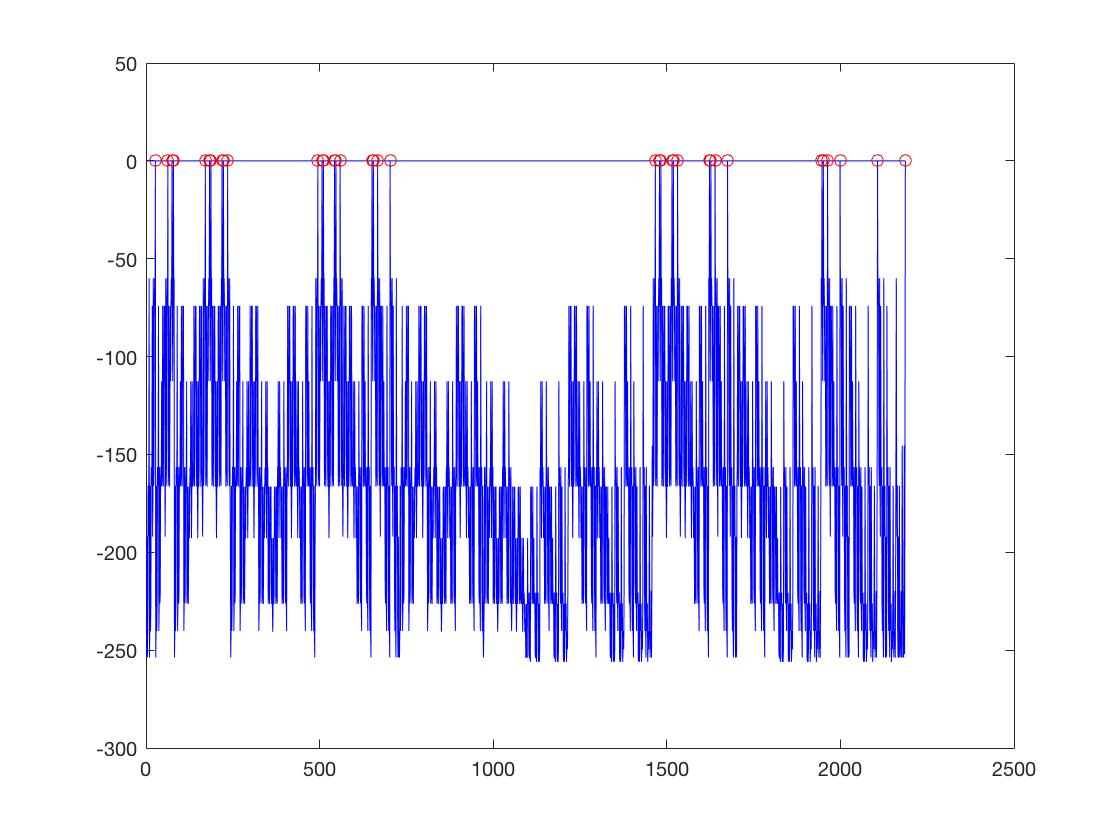}
\end{tabular}
\caption{The left panel is the sensitivity plot of the T-optimal design constructed using a dense grid to show there is no violation in general, and the right panel is the same plot using a sparse grid to show the $36$ support points in total. }
\end{figure}

A limitation of our methodology is that it only applies to polynomial regression models with several factors.  It does not apply to nonlinear models, or even for linear regression models such as fractional polynomials, where the powers in the monomials are certain fractions, or linear models with basis regression functions involving $sine$ and $cosine$. These are useful directions for future research because fractional polynomials, as an example, are increasingly recognized as more flexible than polynomial models and are increasingly used in the biomedical sciences   to model a continuous biological outcome. Another limitation is that our method only returns one optimal design, even when there are several ones, including some designs with smaller number of support points.

\section{Appendix}

We provide here an illustrative MATLAB code for finding the optimal discrimination design in Example 1.

\begin{verbatim}
r = 2; % half degree
mpol x;
K  = [1-x^2 >= 0]; % design space
P  = msdp(K,r);
[F,h,y] = myalmip(P);
M = sdpvar(F(1)); % moment constraint
z = sdpvar(3, 1);
t = sdpvar(1);
t1 = sdpvar(1);
t2 = sdpvar(1);
sol = optimize([F,[M, z; z', t] >= 0, sum(z) <= t1, ...
          max(0, -4*z(1)) + max(0, -4*z(2)) <= t2 ],...
          0.25*t + t1 + t2); % maximize dual problem
ystar = [1; double(y)];
R = msdp( mom(mmon(x, 2*r))==ystar, 1-x^2 >= 0, r+1);
[stat, obj] = msol(R);
\end{verbatim}

\begin{acknowledgements}
All authors gratefully acknowledge partial support from a grant from the National Institute of General Medical Sciences of the National
Institutes of Health under Award Number R01GM107639. The content is solely the responsibility of the authors and does not necessarily
represent the official views of the National Institutes of Health. Dr.\ Lieven Vandenberghe is also partially by a National Science Foundation grant 1509789.
We wish to thank Dr.\ Didier Henrion for helpful discussions and advice
on using the software package GloptiPoly3. We also thank Dr.\ Fabrice Gamboa for his helpful comments on an earlier version of this manuscript.
\end{acknowledgements}

\bibliographystyle{spbasic}      
\bibliography{T-optimal_design}   

%
%

\end{document}